\documentclass[]{aastex631}

\usepackage{float}
\usepackage[inkscapeformat=png]{svg}
\usepackage{xspace}
\usepackage[normalem]{ulem}
\newcommand{\angstrom}{\textup{\AA}\xspace}
\turnoffedit
\turnoffediting
\begin{document}

\title{Investigating the Soft X-ray Spectra of Solar Flare Onsets}

\author[0000-0001-5047-5133]{Anant Telikicherla}
\affiliation{Department of Electrical and Computer Engineering, University of Alberta, 116 St \& 85 Ave, Edmonton, Canada, AB T6G2R3}

\author[0000-0002-2308-6797]{Thomas N. Woods}
\affiliation{Laboratory for Atmospheric and Space Physics, University of Colorado at Boulder, 3665 Discovery Dr., Boulder, CO 80303}

\author[0000-0002-1426-6913]{Bennet D. Schwab}
\affiliation{Laboratory for Atmospheric and Space Physics, University of Colorado at Boulder, 3665 Discovery Dr., Boulder, CO 80303}

\begin{abstract}

In this study we present the analysis of six solar flare events that occurred in 2022, using new data from the third-generation Miniature X-Ray Solar Spectrometer (MinXSS), also known as the Dual-zone Aperture X-ray Solar Spectrometer (DAXSS). The primary focus of this study is on the flare's ``onset phase", which is characterized by elevated soft X-ray emissions even before the flare's impulsive phase. We analyze the temporal evolution of plasma temperature, emission measure, and elemental abundance factors during the flare onset phase, by fitting the DAXSS spectra with the Astrophysical Plasma Emission Code (APEC) model. The model fitting results indicate that the flaring-plasma is already at a high temperature (10-15 MK) during the onset period. The temperature rises during the onset phase, followed by a decrease and subsequent increase during the impulsive phase. Elemental abundance factors show a trend of falling below pre-flare values during the onset phase, with some recovery before the impulsive phase. During the impulsive phase, the abundance factors decrease from elevated coronal values to about photospheric values. We also analyze images from the 193 \angstrom channel of the Atmospheric Imaging Assembly (AIA), highlighting the formation or brightening of coronal loop structures during the onset phase. Two distinct onset loop configurations are observed which are referred to as 1-loop and 2-loop onsets. Both DAXSS and AIA observations indicate that the flare onset phase exhibits similar hot coronal plasma properties as the impulsive phase, suggesting that the onset phase may act as a preconditioning effect for some flares.

\end{abstract}

\keywords{Solar corona (1483); Solar flare spectra (1982); Solar x-ray flares (1816); Solar abundances (1474); Solar x-ray emission(1536)}

\section{Introduction} \label{sec:intro}

The solar Soft X-ray emission spectrum is key diagnostic for understanding the physical phenomena responsible for impulsive plasma heating during solar flares. The soft X-ray spectra is generated from hot coronal plasma (6-16 MK) (e.g., \cite{del_zanna_solar_2018}), and includes both continuum and many emission lines (\cite{fletcher_observational_2011}). Modelling of Soft X-ray spectrum during the course of a flare can reveal the temperature distribution, as well as the variation in the abundance factors of coronal elements. A effective method to understand the sources of coronal heating during flares is observing the change in abundance factors for low first-ionization potential (FIP) elements, i.e., elements with FIP below about 10 eV, such as Si, Ca, and Fe. The abundance change relative to the photospheric abundance is expected to be about two to four for coronal closed magnetic field features and closer to one (photospheric) for open field features \cite{laming_fip_2015}. Furthermore, heating due to magnetic reconnection could show lower elemental abundances in the corona than that from Alfvén dissipation heating due to plasma injection into the coronal loops from the chromosphere \cite{warren_measurements_2014}. Therefore, analyzing the coronal elemental abundance variation during solar flares can reveal the relative importance of different coronal heating mechanisms.

The particular focus of this study is to investigate the flare-onset periods, which exhibit characteristically high solar plasma temperature even before the main flare impulsive period as reported recently by \citep{hudson_hot_2021, battaglia_existence_2023} and references therein. In this study we adopt the definition of the term ‘onset’ described by \citep{hudson_hot_2021} as the “pre-flare interval during which elevated GOES soft X-ray flux is detected, but prior to the detection of any elevated hard X-ray (HXR) emission (at $>$25 keV for stronger events, and 12-25 keV for weaker ones) by the Reuven Ramaty High Energy Solar Spectroscopic Imager (\citep{lin_reuven_updated})”. However, as the RHESSI satellite is no longer operational, no HXR data is available from RHESSI for the solar flares analyzed in this paper that occurred in 2022. Thus, we use the derivative of the Soft X-ray emission as a proxy for the Hard X-ray emissions, which is a consequence of the Neupert Effect \citep{neupert_comparison_1968} that explains the relationship between soft X-rays (SXRs) and hard X-rays (HXRs). Additionally, following \citep{battaglia_existence_2023}, we also emphasize that in this study we analyze the pre-flare onset period and not the “pre-cursor period” which refers to events occurring well before the flare is observed in X-ray energies, such as non-thermal velocity distributions (e.g., \cite{harra_nonthermal_2001}) or SXR precursors  (e.g., \citep{tappin_all_1991}). In this study, we follow the International Organization for Standardization \citep{iso_updated} standard in defining wavelength range, where Soft X-ray wavelengths are defined from 0.124 keV to 12.4 keV and Hard X-ray wavelengths are defined from 12.4 keV to 1240 keV.

We analyze the Soft X-ray spectrum during solar flare onsets, using new measurements from the third generation of NASA-funded Miniature X-ray Solar Spectrometer \citep{woods_first_2023, Woods_2017, MASON20203} instrument, aslo referred to as Dual-zone X-ray Solar Spectrometer (DAXSS). The DAXSS instrument is onboard the INSPIRESat-1 \citep{CHANDRAN20212616} small satellite that successfully launched on February 14, 2022, on the Indian Space Research Organization’s Polar Satellite Launch Vehicle (PSLV-C52).  DAXSS observes the solar Soft X-ray (SXR) spectral irradiance over an energy range of 0.4 keV to 12 keV (0.1 nm to 3 nm). DAXSS has improved energy resolution of 0.05 keV at 1 keV, which is 3 times better than MinXSS-1, and has improved responsivity that is 600 times better than MinXSS-1 (as measured at 4 keV). As described in \citep{Schwab_2020}, the DAXSS instrument was previously flown in 2018 on the NASA 36.336 sounding rocket and took measurements of the quiescent (non-flaring) sun. Now, being on board a satellite in the Low Earth Orbit (LEO), the DAXSS instrument has taken multiple systematic solar soft X-ray measurements, which can be used to analyze the solar coronal plasma during multiple flares in 2022 and 2023. In this study, we present model fitting results from the DAXSS Soft X-ray spectra, to characterize plasma temperature, emission measure, and elemental abundance factors evolution during solar flares onsets. We also analyse Extreme Ultaviolet (EUV) images using the Solar Dynamics Observatory (SDO) Atmospheric Imager Assembly (AIA) \citep{aia} instrument, to reveal the spatial variation of coronal plasma structures during solar flare onsets. By modelling DAXSS Soft X-ray spectra and analysing SDO AIA EUV images during flares together, in this  study we aim to answer the following specific questions: 

\begin{enumerate} 
    \item What is the variation in coronal Plasma Temperature \& Emission Measure during flare onset and impulsive phases?
    \item What is the trend in Elemental Abundance ratio variation of low First Ionization Potential (FIP) elements during the flare onset and impulsive phases? 
    \item What is the spatial origin and morphology variation of the coronal plasma during the flare onset period?
\end{enumerate}

Overall, these questions aim to reveal the physical processes leading to impulsive energy release during Flare Onset and Impulsive phases. The paper first describes the spectral fitting methodology adopted for this analysis, including the type of model used and the energy range for fitting. This is followed by spectral fitting results of six different solar flares during 2022. Then, analysis of SDO AIA EUV images are presented, highlighting the spatial variation of plasma structures. The paper concludes by summarizing the common trends observed in the six flares, and possible implications and inferences.

\section{Spectral Fitting Methodology} \label{sec:models}

\begin{figure}
\centering
\includegraphics[width = \textwidth]{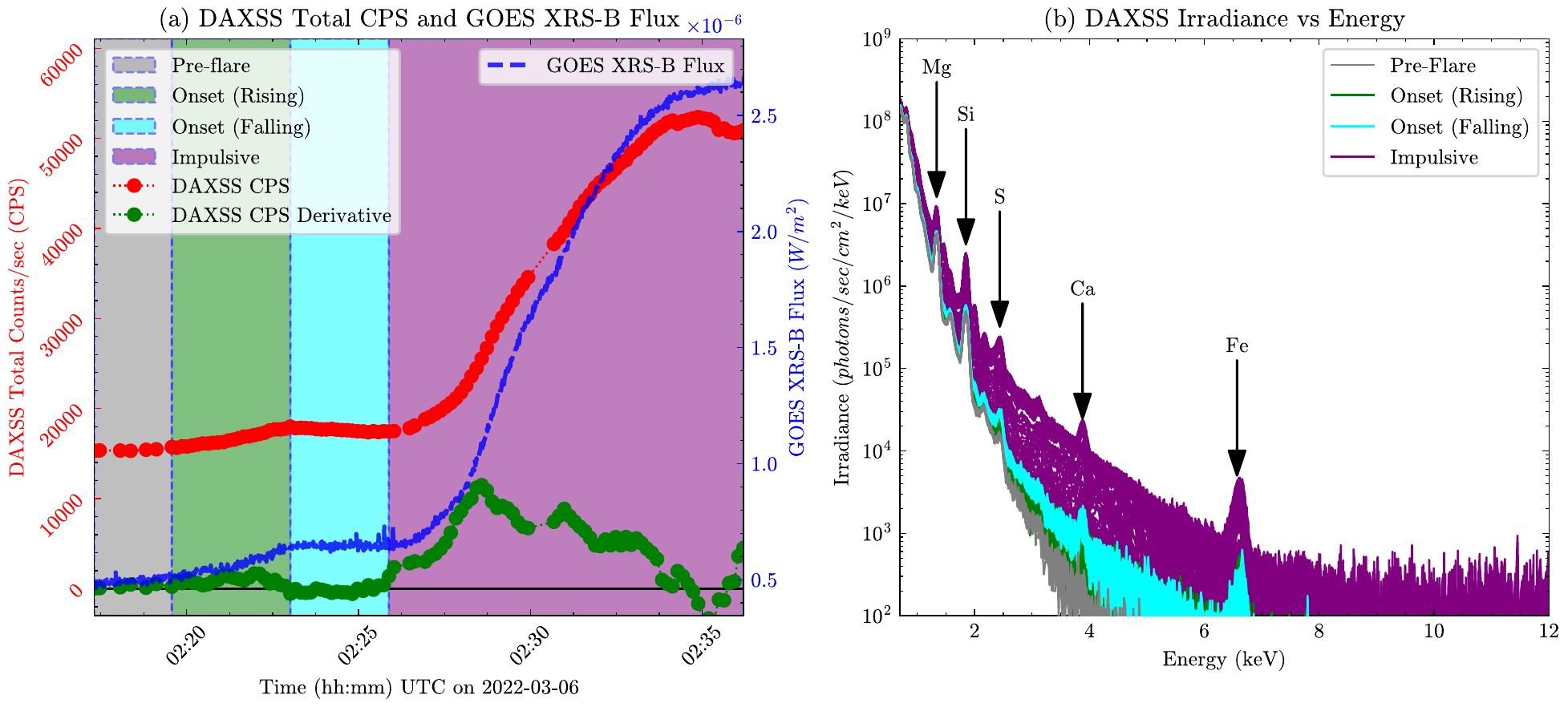}
\caption{Panel (a) shows DAXSS Total Counts per second (CPS) (red) and GOES XRS-B Flux (blue) during a C2.5 flare on 2022-03-06. The phases of the flare are indicated with different color shades: Pre-flare (grey), Onset-Rising (green), Onset-Falling (blue) and Impulsive (purple). The derivative of the DAXSS total counts per second is shown as green dots. The derivative has been smoothed and scaled to fit into the same axis scale. During the onset phase, the derivative is close to zero or negative. The onset phase ends when the derivative value changes from negative to positive, indicating an increase in the Hard X-ray flux. Panel (b) depicts the evolution of measured energy spectra during the different phases of the flare. Line complexes for five elements: Si, Ca, Fe, Mg, and S are marked using black arrows.}
\label{doy_074_cps_spectra}
\end{figure}

The spectral fitting in this study is performed using PyXSPEC, which is the Python interface to the XSPEC \citep{arnaud_xspec_1996} spectral-fitting program. The DAXSS Level-1 data product contains the spectra (counts/second for the various energy bins of the instrument) which are converted to Pulse Height Analyzer (PHA) files for analysis in PyXSPEC. Response Matrix File (RMF) and Auxiliary Response File (ARF) were also generated, which contain information about the energy redistribution and effective area of the instrument respectively. The process of fitting in XSPEC is started by selecting an appropriate model (described in the subsequent section), convolving the model with the instrument response (RMF and ARF files), and comparing this model to the instrument measurements using reduced chi-squared ($\chi^2$) as a goodness-of-fit indicator. The model parameters such as plasma temperature, emission measure and elemental abundances are varied to get the minimum reduced-$\chi^2$. This process is then repeated for all measurements during the solar flare to obtain the temporal evolution of the fit parameters. As an example, Figure \ref{doy_074_cps_spectra}(a) shows the total measured X-ray photon counts (red, overlayed with the GOES XRS-B flux (blue)) of a C2.5 class flare on 2022-03-06. Each flare analyzed in this paper is divided into pre-flare and flare components. The flaring component is further divided into onset (which consists of onset-rising and onset-falling), impulsive and gradual phases to aid the discussions. Figure \ref{doy_074_cps_spectra}(b) shows the evolution of the irradiance over the 0.7 to 12 keV energy range for the pre-flare (grey), onset-rising (green), and onset-falling (blue), and impulsive (purple) phases of the flare. The various line complexes in the soft X-ray range, including Mg, Si, S, Ca, and Fe are also shown in figure \ref{doy_074_cps_spectra}(b). Further details of the various line complexes in the DAXSS spectra are described in \cite{woods_first_2023}.

\subsection{Energy Range for Spectral Fitting}
DAXSS has a measurement range from 0.4 keV to 12 keV, however, the instrument response is yet to be validated between 0.4 keV and 0.7 keV, thus 0.7 keV is chosen as the lower energy limit for this analysis. For the higher energy limit, three options are considered: (a) Fixed higher limit at 4.0 keV, (b) variable high energy limit according to the counts of the spectra (i.e. energy limit is where the counts are less than 2 per integration time (which is 9 seconds)), and (c) fixed higher energy limit at 7.0 keV. As shown in figure \ref{doy_074_cps_spectra}(b), the photon counts measured at the different energy bins vary with the measurement time (for example pre-flare vs onset vs flare-peak), thus choosing a fixed higher-energy limit of 7.0 keV leads to bad fits because of fewer counts in the range of 4.0 to 7.0 keV for the pre-flare spectra. Similarly, fixing a limit of 4.0 leads to not being able to fit the higher energy Fe line during the flare spectra. Hence, an approach of dynamically selecting the upper energy limit based on counts was adopted for this analysis. In this method, the number of counts in each bin is iteratively checked until the counts become less than a certain threshold (2 counts per integration time). The energy bin where the counts become less than the threshold is considered the upper limit. 

\subsection{Model Selection for Spectral Fitting}
\begin{figure}
\centering
\includegraphics[width = 0.49\textwidth]{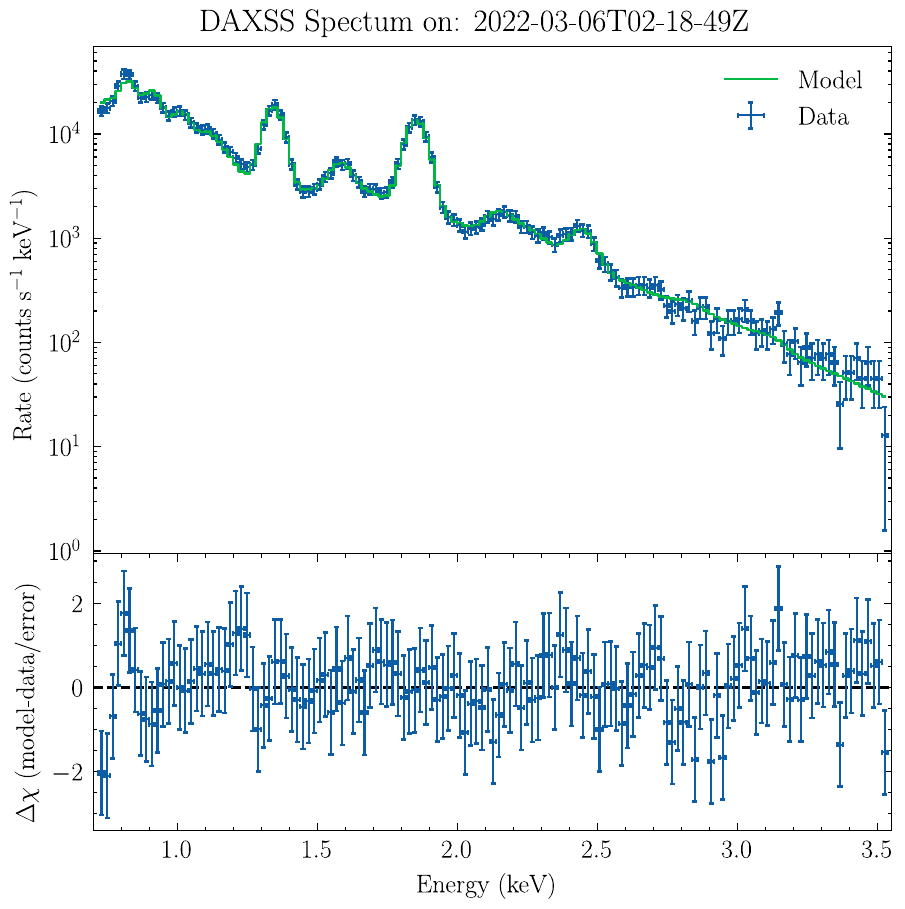}\hfill
\includegraphics[width = 0.49\textwidth]{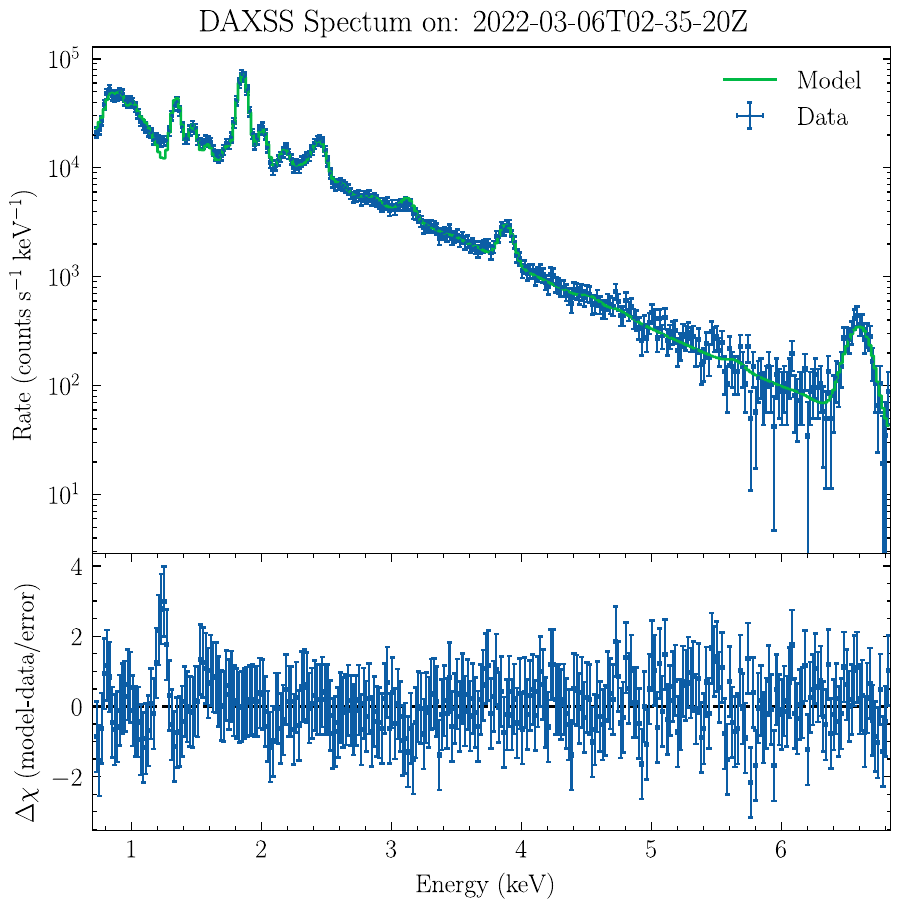}
\caption{Model fit results for two particular spectra of a C2.5 flare on 2022-03-06. The left panel shows the pre-flare spectra model fit and right panel shows the impulsive phase spectra model fit. The measured count rates (blue, with error bars) are fit to a VVAPEC model spectrum (green). The bottom panel shows $\Delta \chi$ (model-measurement/error) for each measurement, indicating the closeness of the model prediction to actual measurements}
\label{doy_074_spectra_fits}
\end{figure}

The Astrophysical Plasma Emission Code (APEC) model used for the analysis in this paper, uses data from the AtomDB Atomic Database \citep{Foster_2012} to calculate spectral models of thermal emission from optically thin plasma. The APEC model (particularly VVAPEC), consists of various parameters including Plasma Temperature, Emission Measure, and Elemental Abundance Factor Ratios which can be varied during model fitting. For this analysis, abundances are based on the Feldman standard extended coronal (FSEC) abundance values \citep{1992ApJS...81..387F, Landi_2002}. Based on the dynamically selected upper energy range, different elemental abundance factors are included in the model. Typically these include low first-ionization potential (FIP) elements like Si, Ca, Fe, Mg, and S. Since solar flares generally consist of multiple plasma-temperature components, models with multi-temperature plasma components are used. Further, the relative abundance of the 5 elements (Si, Mg, S, Ca, and Fe) can also be independently varied for each of the model components. These options thus enable the possibility of having a wide variety of models that can be considered for fitting. \cite{Nagasawa_2022} present a comparison of fitting with 1T, 2T, and 3T VVAPEC models with the elemental abundance fixed. Based on the procedure for model fitting described in \cite{woods_first_2023} and \cite{Nagasawa_2022}, for this study, we use a 3-Temperature Model in which the elemental abundances (of Si, Ca, Fe, Mg, and S) corresponding to the three temperature components are tied together. Thus the model consists of 11 variables (3 Temperatures, 3 Emission Measures, and 5 Elemental Abundances). As an example, Figure \ref{doy_074_spectra_fits} shows the model fit for the pre-flare spectra and one of the spectra during the impulsive phase of a C2.5 flare on 2022-03-06. The impulsive phase spectrum in figure \ref{doy_074_spectra_fits} (b), shows a overall enhanced count rate in the energy range and subsequently has a higher energy upper limit (approx. 7 keV) as compared to the pre-flare spectra which has a lower upper limit (approx. 3.5 keV).

\section{Flare Soft X-ray Fitting Results}
Table \ref{tab:flarelist} presents a list of six flares that have been analyzed in this study, along with the Day of Year (DOY), UTC date and soft X-ray flux peak time of each flare. Each flare is assigned a number from 1-6, to aid the discussion. The table also lists the Geostationary Operational Environmental Satellite (GOES) Soft X-ray Flux classification, and the six flares are chosen from the DAXSS measurements, cover a wide range of GOES classes from C2.5 to M2.7. The type of flare is also listed which can be (a) eruptive, where plasma is ejected from the solar atmosphere resulting in a Coronal Mass Ejection (CME)) or (b) confined, which is when no CME is associated with the flare. The flares also have different spatial locations, ranging from near the center as well as towards the limb, which is indicted by the helio-projective co-ordinates of the flaring region. The table also lists the bottom-left and top-right helio-projective co-ordinates of a bounding box around the flaring region. These co-ordinates are used to generate AIA cutouts in the subsequent sections. They six flares also have different different duration's and temporal profiles, with the impulsive phases of some being steeper than others. Overall, these six flares represent various different types of flares, to ensure that the results are not biased only for one particular type of flare.  

\begin{figure}
\centering
\includegraphics[width = \textwidth]{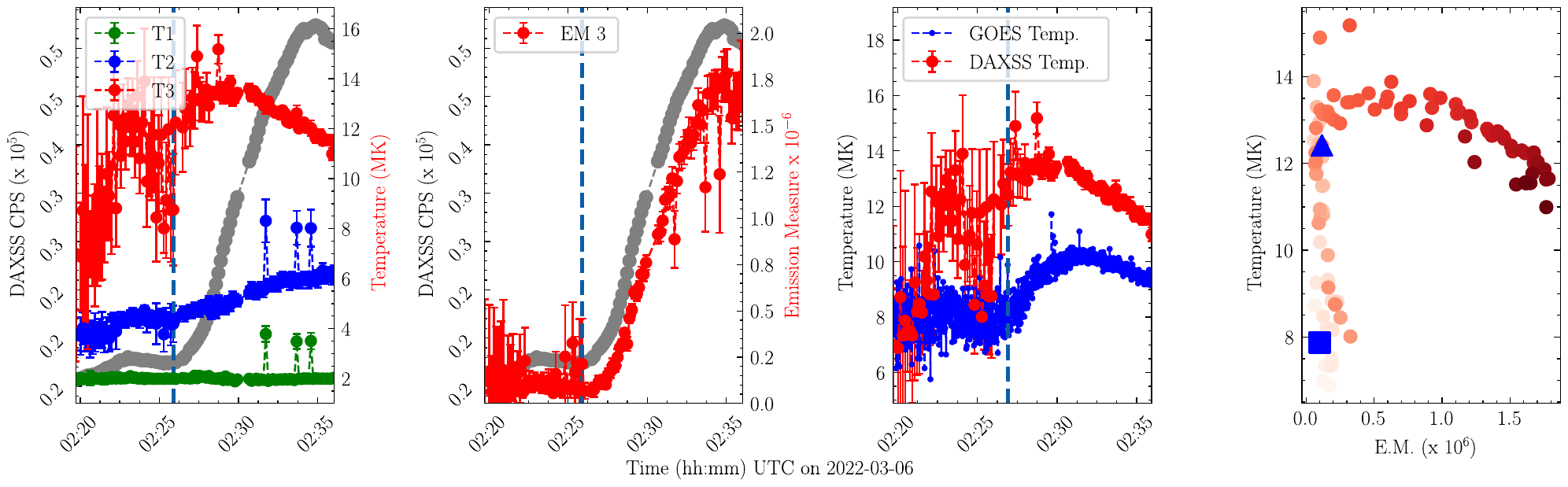}
\includegraphics[width = \textwidth]{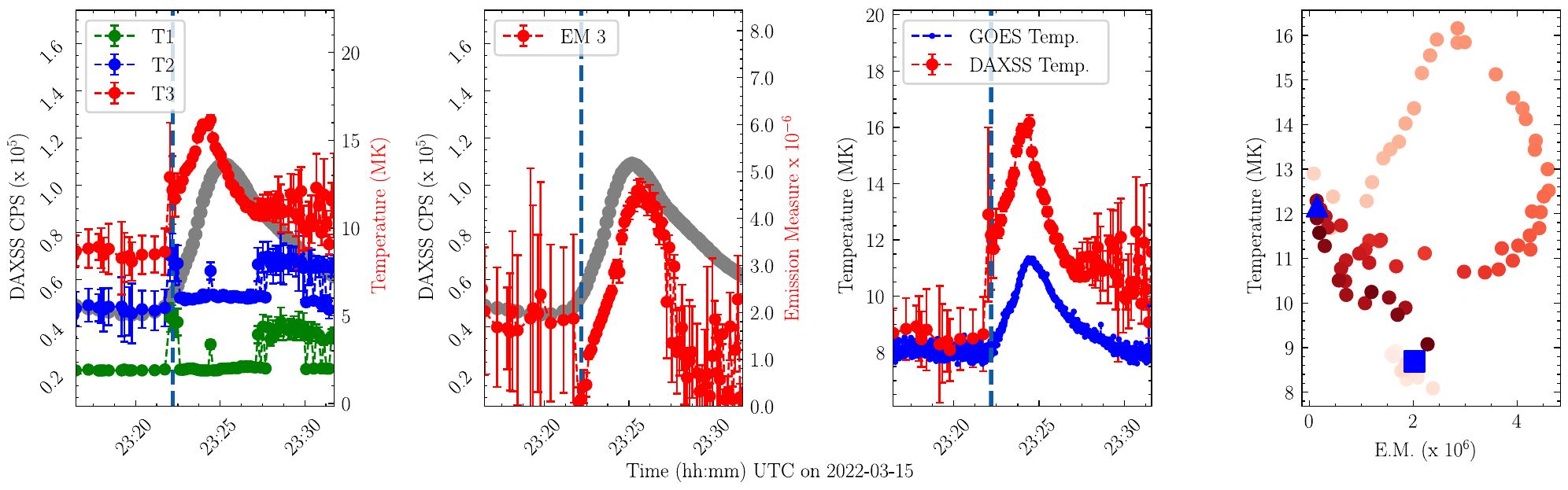}
\includegraphics[width = \textwidth]{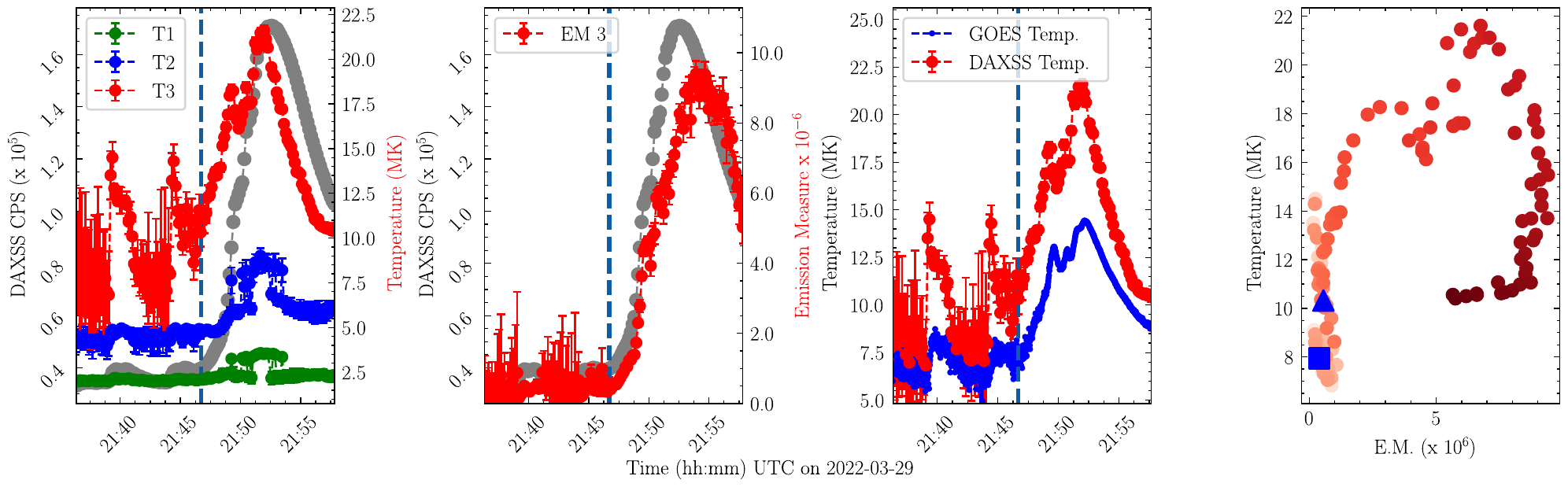}
\caption{Temperature and emission measure model fit results for flares 1-3: First column shows the Plasma Temperature variation (three components indicated in red, blue and green), overlayed with total measured counts per second (grey). Second Column shows plasma emission measure variation, of the hottest plasma temperature component. The third column shows a comparison of the hottest DAXSS temperature (red) with the GOES isothermal temperature (blue). The vertical dashed line in the first three columns indicates the end of the onset phase of the flare. The fourth column shows the hottest temperature vs emission measure, the colour gradient (light to dark) indicates increasing time. In the fourth column, the first data point (start of pre-flare period) is marked using the square symbol and the end of the onset phase is marked using a triangle symbol.}
\label{results_T_EM_1}
\end{figure}

\begin{figure}
\centering
\includegraphics[width = \textwidth]{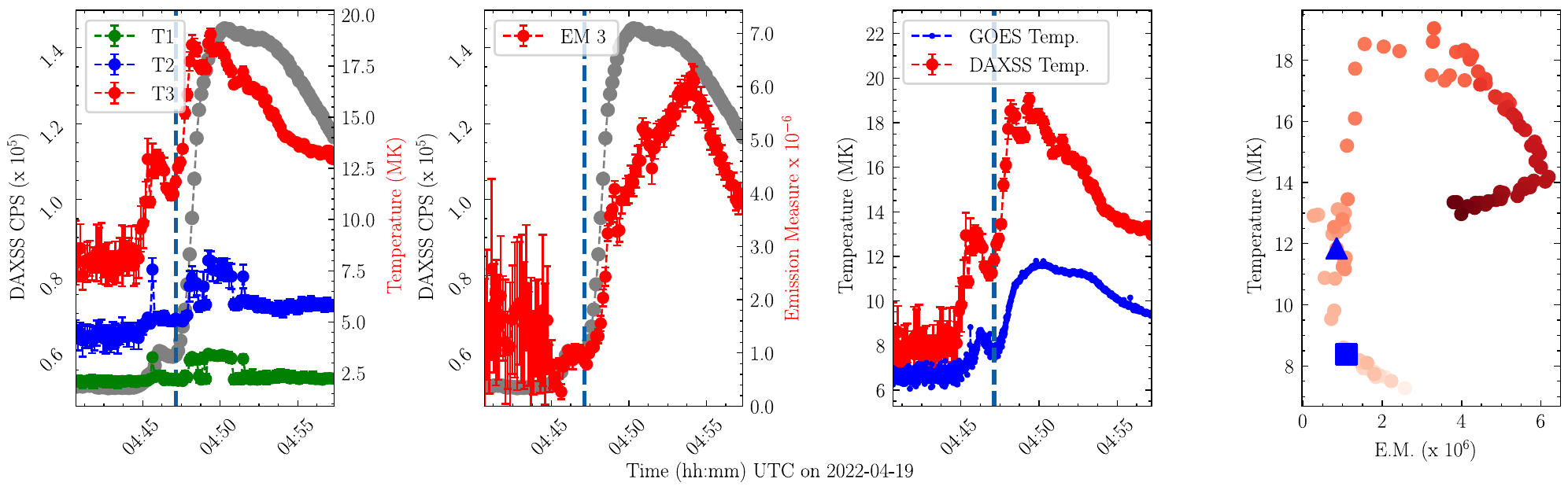}
\includegraphics[width = \textwidth]{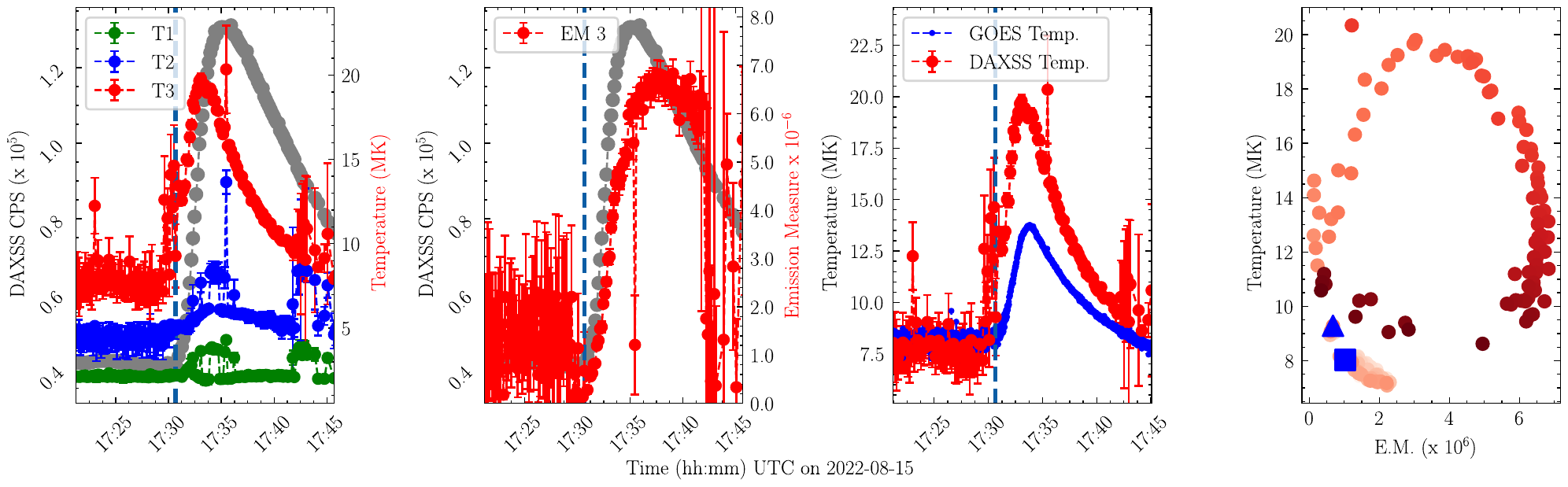}
\includegraphics[width = \textwidth]{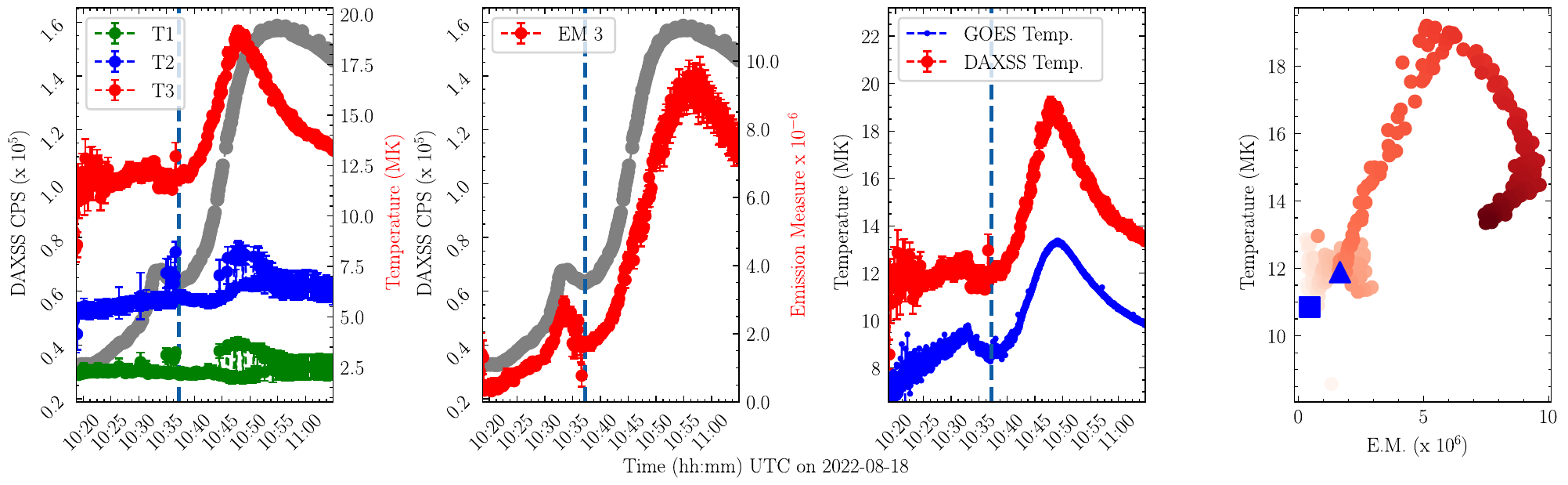}
\caption{Temperature and emission measure model fit results for flares 4-6: First column shows the Plasma Temperature variation (three components indicated in red, blue and green), overlayed with total measured counts per second (grey). Second Column shows plasma emission measure variation, of the hottest plasma temperature component. The third column shows a comparison of the hottest DAXSS temperature (red) with the GOES isothermal temperature (blue). The vertical dashed line in the first three columns indicates the end of the onset phase of the flare. The fourth column shows the hottest temperature vs emission measure, the colour gradient (light to dark) indicates increasing time.In the fourth column, the first data point (start of pre-flare period) is marked using the square symbol and the end of the onset phase is marked using a triangle symbol.}
\label{results_T_EM_2}
\end{figure}

\setlength{\LTcapwidth}{\textwidth}
\begin{center}
\begin{longtable}{ | m{0.8cm} | m{2cm}| m{4cm} | m{1.5cm} | m{1cm} | m{2.5cm} | m{2.5cm} | } 
\hline
\textbf{S.No.} & \textbf{Flare Number \& DOY} & \textbf{Date-Time (UTC) of GOES XRS Peak} & \textbf{ Type} & \textbf{Class} & \textbf{Bottom-Left Co-ordinate} & \textbf{Top-Right Co-ordinate} \\ \hline
1              & Flare-1 (DOY-065)     & 2022-03-06, 02:37    &  Confined      & C2.5             
& (-750, 450) & (-650, 600)\\ \hline

2              & Flare-2 (DOY-074)     & 2022-03-15, 23:25     & Eruptive       & C6.7 
& (100, 400) & (400, 600)\\ \hline

3              & Flare-3 (DOY-088)     & 2022-03-29, 21:52    &  Confined      & M1.6             
& (-300, 300) & (500, 450)\\ \hline
4              & Flare-4 (DOY-109)     & 2022-04-19, 04:50    &  Confined      & M1.0             
& (50, -250) & (200, 1-50)\\ \hline
5              & Flare-5 (DOY-227)     & 2022-08-15, 17:00    &  Confined      & M2.7           
& (550, -400) & (900, -100)\\ \hline
6              & Flare-6 (DOY-230)     & 2022-08-18, 10:55   &  Confined      & M1.5             
& (400, -600) & (650, -400)\\ \hline
\caption{List of Flares Analyzed with the flare number, Day of Year (DOY), UTC Time of GOES XRS Peak, Flare Type (Confined or Eruptive), GOES X-ray Flare Class, and helioprojective co-ordinates of the flaring region}
\label{tab:flarelist}
\end{longtable}
\end{center}
Figures \ref{results_T_EM_1} and \ref{results_T_EM_2}, show the model fitting results from all six flares following the fitting procedures explained in the earlier section. The first column shows the plasma temperature (in MK) obtained from through the evolution of the flare (the x-axis is UTC time). The three temperatures are color-coded, Temperature 1 (green - Coolest Temperature), Temperature 2 (blue) and Temperature 3 (red, Hottest Temperature). Similarly, the second column shows the XSPEC norm corresponding hottest plasma temperature component (Temperature-3). The emission measure corresponding to the lower temperatures (1 and 2) are not plotted since they are larger in magnitude, and mask the trend of the emission measure corresponding to the hottest temperature (which is the main focus of this study).  The XSPEC norm is referred to in this paper as emission measure for simplicity. The mathematical relation between the XSPEC norm and plasma emission measure can be found in \citep{leahy_interpretation_2023}. In both the temperature and emission measure plots, the DAXSS total counts per second is shown in Grey, to indicate the evolution of the solar soft X-ray emission. The error bars in these plots are obtained from the XSPEC model parameter sigma value, and indicate the one sigma confidence intervals (details of their calculation using the second derivatives of the fit statistic with respect to the model parameters at the best-fit can be found in \citep{arnaud_xspec_1996}). The third column shows a comparison of the hottest DAXSS temperature (Temperature 3) with the GOES isothermal temperature (blue). The GOES isothermal temperatures are obtained using the GOES-16 X-ray Sensor (XRS) A/B ratio method, described in detail in \citep{woods_first_2023}. In the first three columns a dashed blue line is drawn to indicate the end of the onset phase and beginning of the main flare impulsive phase. The fourth column gives a plot of the temperature vs emission measure during the flare. This plot has a red color gradient depicting time, with lighter colors representing the start of the flare and darker colors being towards the end of the flare. The first data point (start of the pre-flare) is indicated using a blue square symbol and the end of the onset phase is indicated using a blue triangle symbol in the temperature vs emission measure plot.

In all six flare events, it is observed that the localized plasma temperature rises during the impulsive phase, peaks before the peak in the Soft X-ray emissions, and then slowly falls back to the original value during the gradual phase (Column 1 of Figures \ref{results_T_EM_1} and \ref{results_T_EM_2}). For all six flares, the plasma temperatures are observed to already rise to a higher value in the range of 10-15 MK during the onset phase. This is followed by a decrease in temperature, and then rise again during the main impulsive phase of the flare. The isothermal temperature fit results from GOES X-ray flux data also indicate a similar trend as the hottest temperature obtained from DAXSS (Column 4 of Figures \ref{results_T_EM_1} and \ref{results_T_EM_2}). The emission measures of the three plasma components indicate that increase during the onset and impulsive phase, followed by recovery to original values during the gradual phase (Column 2 of Figures \ref{results_T_EM_1} and \ref{results_T_EM_2}). Column 3 shows the correlation between the hottest Temperature (Temp-3) and Emission Measure (Flare EM 3) for all the flares. The colour gradient of the plot corresponds to the measurement time, with lighter red shades corresponding to the beginning of the flare-onset, and darker shades corresponding to the end of the flare. The plots indicate a clockwise light-to-dark trend, depicting heating during the main flare impulsive phase and cooling during the gradual phase. For some flares, there is a significant increase (2-4 MK) in the plasma temperature with very less increase in the plasma emission measure during the onset phase (as denoted by the locations of the the square and triangle symbols in columns four of figures \ref{results_T_EM_1} and \ref{results_T_EM_2}). The plots also indicate that during the onset phase, even at lower emission measures, the plasma temperatures are already in the 10-15 MK range.

\begin{figure}
\centering
\includegraphics[width = \textwidth]{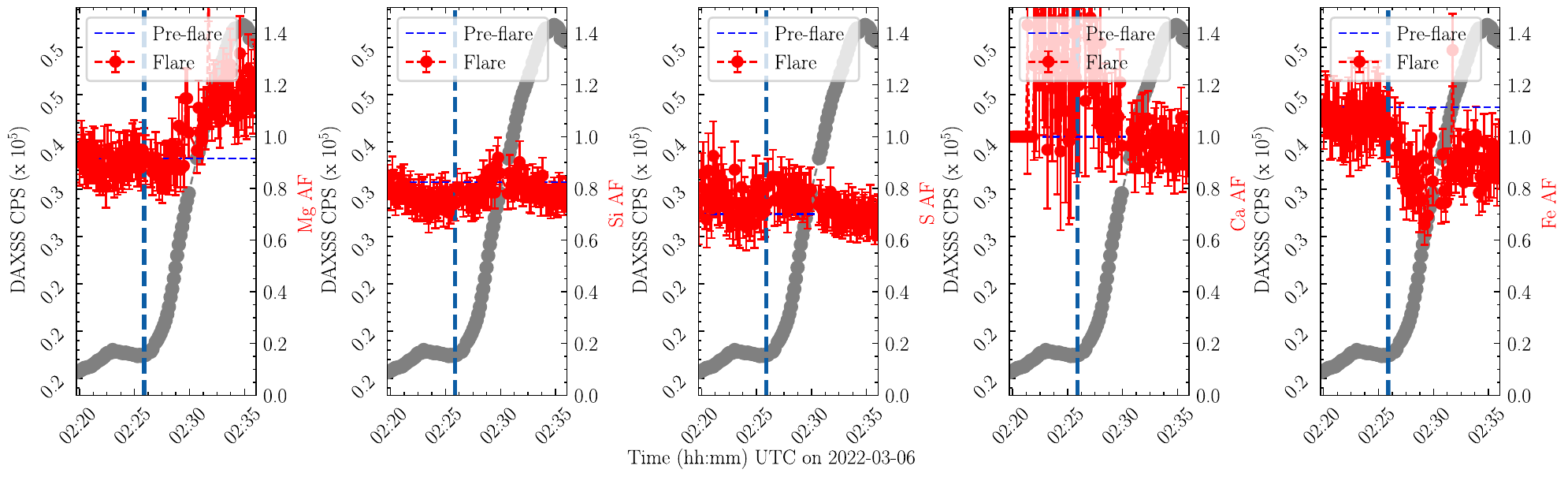}
\includegraphics[width = \textwidth]{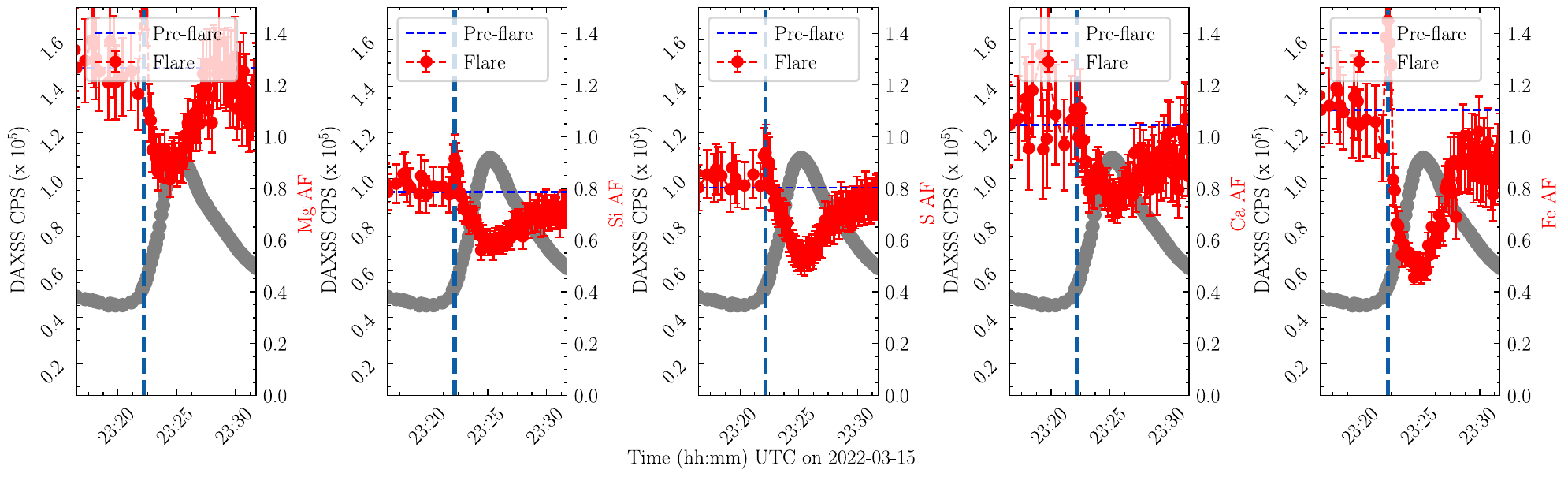}
\includegraphics[width = \textwidth]{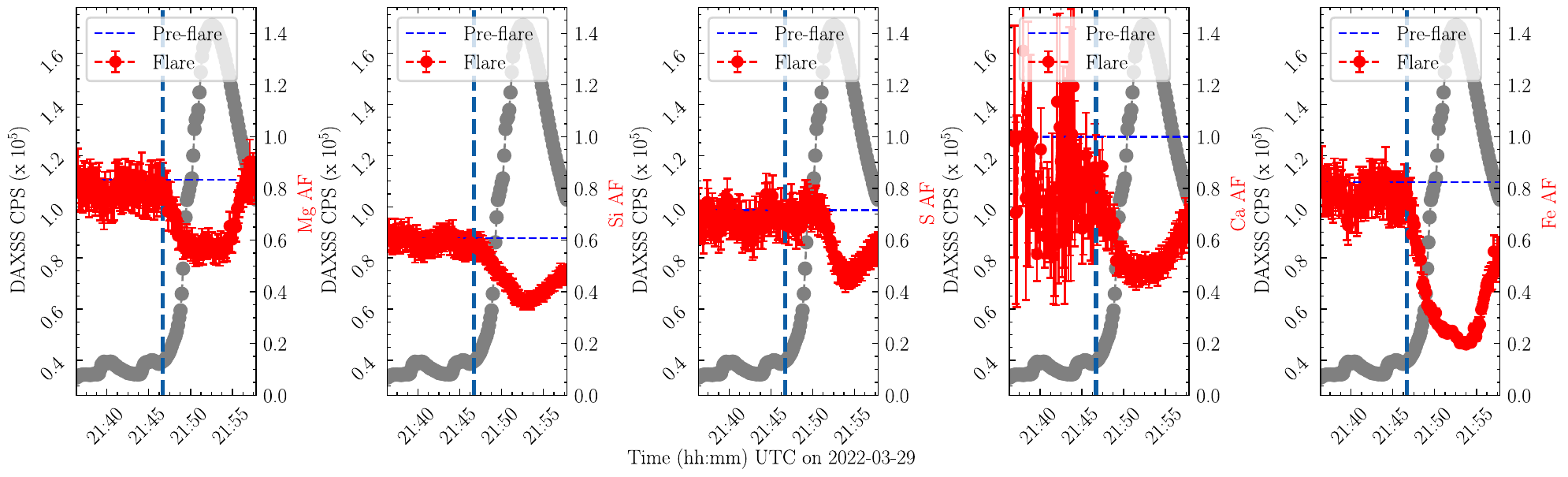}
\caption{Elemental Abundance Factors fit results for flares 1-3: The abundance factors variation for Mg, Si, S, Ca, and Fe are shown in red. The dotted blue lines indicate the pre-flare values. An Abundance Factor of 1 implies the abundance is equal to Feldman standard extended coronal (FSEC) value. The abundance factor plots are overlayed on the observed total Counts Per Second (CPS) shown in Grey, indicating the evolution of the flare. The vertical dashed line in all the columns indicates the end of the onset phase of the flare. Larger error-bars are observed for Ca, primarily during the pre-flare period. This can be attributed to lesser counts during these times at this energy (around 4 keV), which leads to a comparatively weaker fit, with higher uncertainties.}
\label{results_AF_1}
\end{figure}

\begin{figure}
\centering
\includegraphics[width = \textwidth]{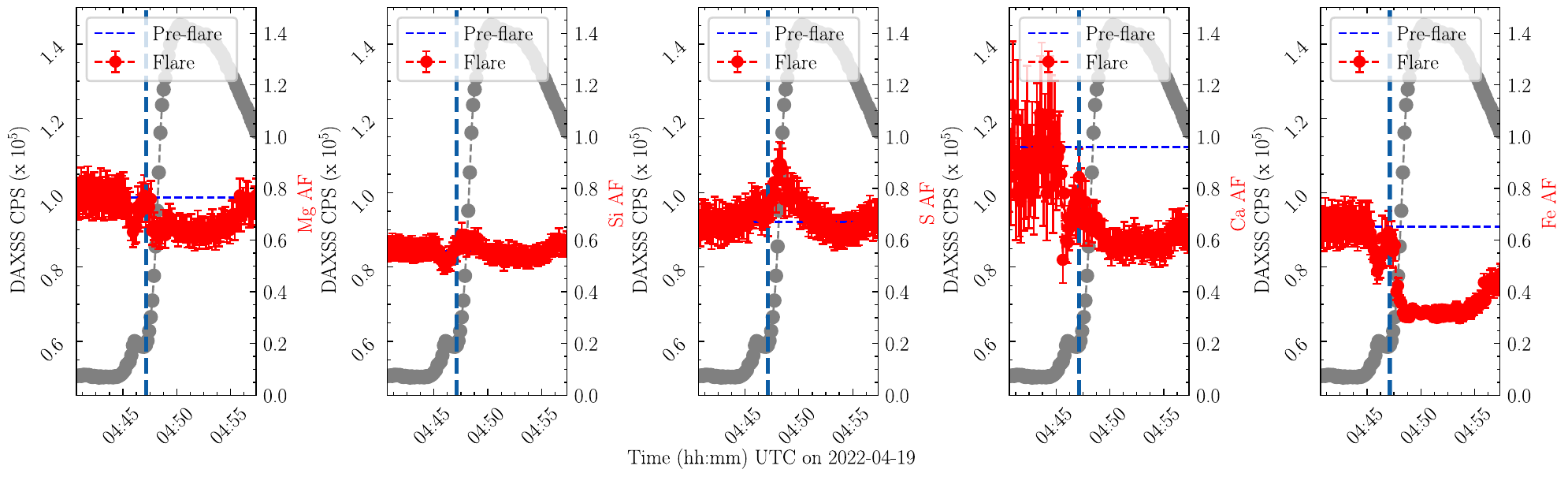}
\includegraphics[width = \textwidth]{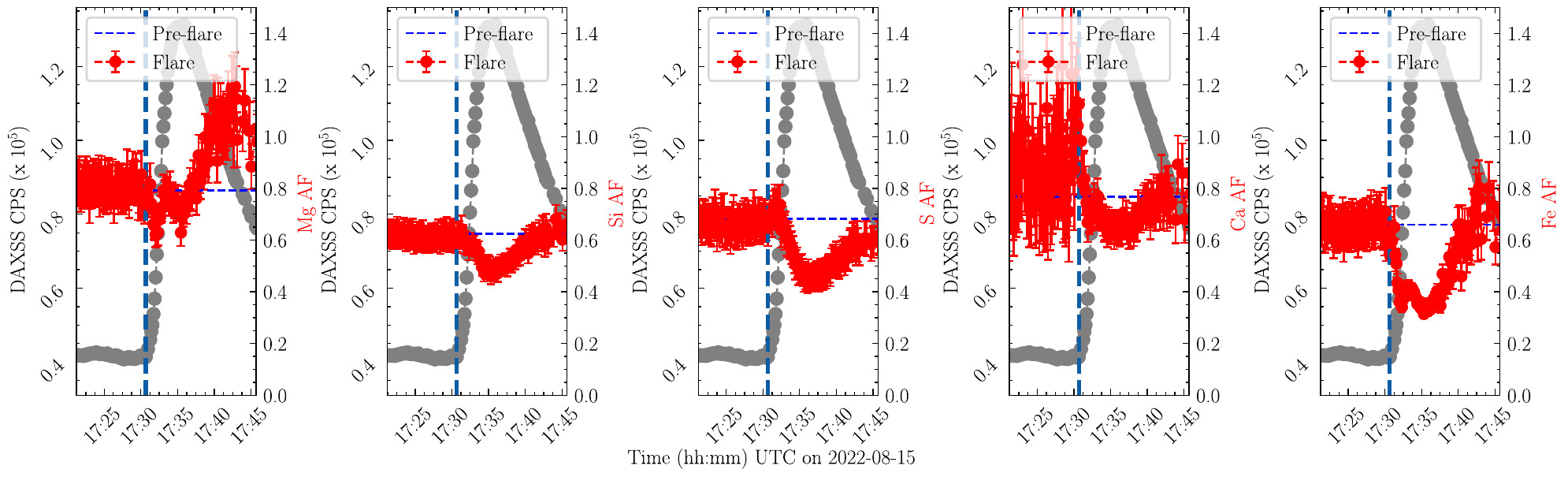}
\includegraphics[width = \textwidth]{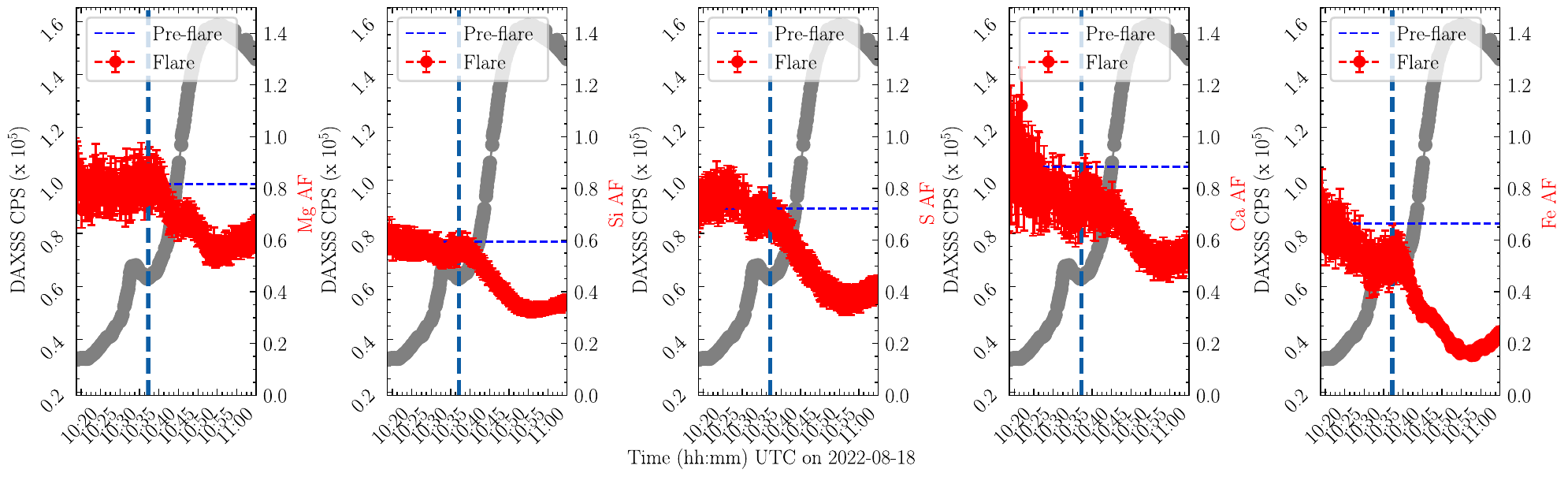}
\caption{Elemental Abundance Factors fit results for flares 4-6: The abundance factors variation for Mg, Si, S, Ca, and Fe are shown in red. The dotted blue lines indicate the pre-flare values. An Abundance Factor of 1 implies the abundance is equal to Feldman standard extended coronal (FSEC) value. The abundance factor plots are overlayed on the observed total Counts Per Second (CPS) shown in Grey, indicating the evolution of the flare. The vertical dashed line in all the columns indicates the end of the onset phase of the flare.}
\label{results_AF_2}
\end{figure}

Figures \ref{results_AF_1} and \ref{results_AF_2}, show the abundance factor variations of Mg, Si, S, Ca, and Fe for the six flares. The dashed-blue lines in these plots indicate the pre-flare abundance factor value and the red lines with error-bars indicate the abundance factor value variation during the flares. These plots are expressed in terms of the Feldman standard extended coronal (FSEC) abundance values, i.e. a Abundance Factor of 1 indicates the standard coronal abundance. Similar to the plasma temperature and emission measure, the errorbars of the abundance factors are obtained from XSPEC and indicate the one sigma confidence interval of the best fit parameters. Larger error-bars are observed for Ca, primarily during the pre-flare period. This can be attributed to lesser counts during these times at this energy (around 4 keV), which leads to a
comparatively weaker fit, with higher uncertainties. All five elemental abundances follow a similar trend, they first decrease and reach a minimum at approximately the same time when the temperature is maximum. The abundance factors then recover to the original value during the gradual phase of the flare. However, the magnitude of the decrease is different for different types of flares. In all abundance factor plots, a dashed blue line is drawn to indicate the end of the onset phase and beginning of the main flare impulsive phase. In some flares, during the onset-rising phase, the abundance factors show a trend of falling to lower values, and recovery towards original values during the onset-falling phase (i.e. a decrease and increase is observed before the main flare impulsive phase.)

\section{Flare EUV Image Analysis}

To understand the spatial location of the flare onset coronal emissions, as well as the plasma morphology variations during the onset and impulsive phases, the Solar Dynamics Observatory (SDO) Atmospheric Imager (AIA) Assembly Extreme Ultra-Violet (EUV) images are analyzed. Among the various wavelengths of the AIA, the 193 \angstrom channel is chosen. This spectral line is emitted by iron atoms that have lost 11 electrons (also known as iron-12 or Fe XII) at formation temperatures near 1,000,000 K (or 1 MK) as well as iron atoms that have lost 23 electrons (also known as iron-24 or FeXXIV) at temperatures of 20 MK (\citep{lemen_atmospheric_2012}). The former represents a slightly higher region of the corona and the latter represents the much hotter material of a solar flare. These temperatures are very close to the temperatures obtained from DAXSS and GOES Soft X-ray data, hence are chosen to investigate the spatial origins and dynamics of the coronal plasma during the onset and impulsive phases. Cutouts of the 193 \angstrom channel images are generated using SunPy \citep{community_sunpy_2020}, according to the helio-projective coordinates listed in Table \ref{tab:flarelist}. The images are also corrected for the change in exposure time, noting that the exposure time changes during the impulsive phase of the flare. To clearly identify the movement of the coronal plasma, running-difference images are generated by subtracting consecutive images. This results in a black and white gradient picture with white indicating positive change, and black indicating negative change in flux. Thus, patterns of black-white features indicate the movement of plasma from black to white in consecutive images. The six flares analyzed are observed to fall under two onset loop-configuration categories.  

The first category is when a small coronal loop emerges or brightens during flare onset that then rises and appears to interact with a higher, sometimes bigger, loop which then brightens during the main flare impulsive phase.  Thus, we refer to this category as 1-loop onset flares because only one loop dominates at a time. Based on the AIA 193 \angstrom\ images Flares 1 and 4 fall under this category.  The second category is when two small loops emerge or brighten during flare onset and merge together to form a bigger, brighter loop (and sometimes eruption) during the main flare impulsive phase. Consecutively, we refer to these as 2-loop onset flares, because two loops are brighter during the onset period. Based on the AIA 193 \angstrom images Flares 2, 3, 5, and 6 fall under this category. A representative example of both categories are described in the subsequent subsections. The flare animations generated of all flares can be found in the supplementary material of the paper. For each flare both the standard AIA 193 \angstrom animations, as well as the running difference animations are included. The format of the file names for the AIA 193 \angstrom images is `DOY\_NNN\_Col.mp4', where NNN is denotes the Day of Year (for e.g., DOY\_065\_Col.mp4) and the suffix `Col' denotes `Colored'. Similarly the format of the file names for the difference images is `DOY\_NNN\_Diff.mp4' where the suffix `Diff' denotes `Difference Image'. The timestamp of each image in the animations is also shown on the top of the image. These animations are for the entire duration of the flares, however the figures in the subsequent subsections show four still frames from two of the flare animations highlighting the key features of the onset and main flare impulsive phases of the flare.

\subsection{1-Loop Onset Flare Example: Flare-1, C2.6 flare on 2022-03-06}
\begin{figure}
\centering
\includegraphics[width = \textwidth]{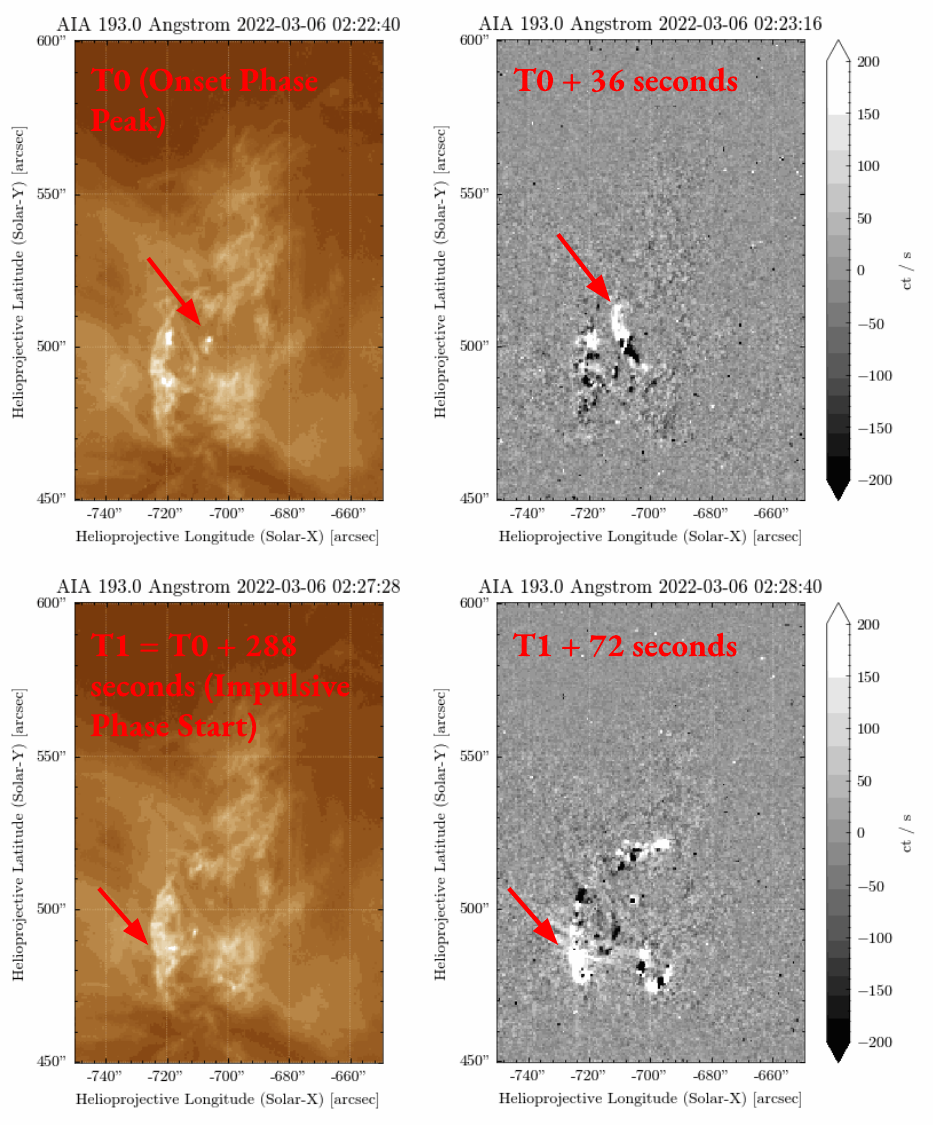} 
\caption{AIA 193 \angstrom Images for a 1-loop onset Category Example: Flare-1, C2.6 flare on 2022-03-06. The first row show images during the flare onset phase and the second row during the flare impulsive phase. The first image on the first row is at T0 (02:22:40 UTC) which is during the Soft X-ray peak of the Onset phase. The next image is difference image, and the time interval of the difference image is shown in red. The first image of the second row is at T1 (02:27:28 UTC) which is during the start of the impulsive phase. Similar to the first row, the next image is a difference image, with the time interval labelled.}
\label{doy_066_aia}
\end{figure}

Figure \ref{doy_066_aia} shows SDO AIA 193 \angstrom channel images of a C2.6 flare on 2022-03-06. The top-left panel shows a cutout during the time of Soft X-ray peak of the onset phase (Labelled as T0). The next image is a running difference image with the time interval after T0 labelled. The first image in the second row shows the AIA 193 \angstrom image at the beginning of the impulsive phase, denoted as T1. Similar to the first row image, the subsequent image is a running difference image with the time interval after T1 labelled. From the difference images, it is observed that a plasma moves in a small loop-like structure (indicated by a black-to-white change in images) during the Onset Phase. This is depicted in the images in the first row. Then during the start of the impulsive phase, the main emissions (brightening) are from plasma at the end point of this loop. These AIA images in the second row indicate that a different plasma loop is responsible for the onset emissions, even before the impulsive phase. This then triggers the main impulsive phase with stronger emissions in the plasma surrounding the end point of the small loop. The supplementary material includes animations for both the AIA 193 \angstrom images (DOY\_065\_Col.mp4) and  the running difference images (DOY\_065\_Diff.mp4).

\subsection{2-Loop Onset Flare Example: Flare-3, M1.6 flare on 2022-03-29}
\begin{figure}
\centering
\includegraphics[width = \textwidth]{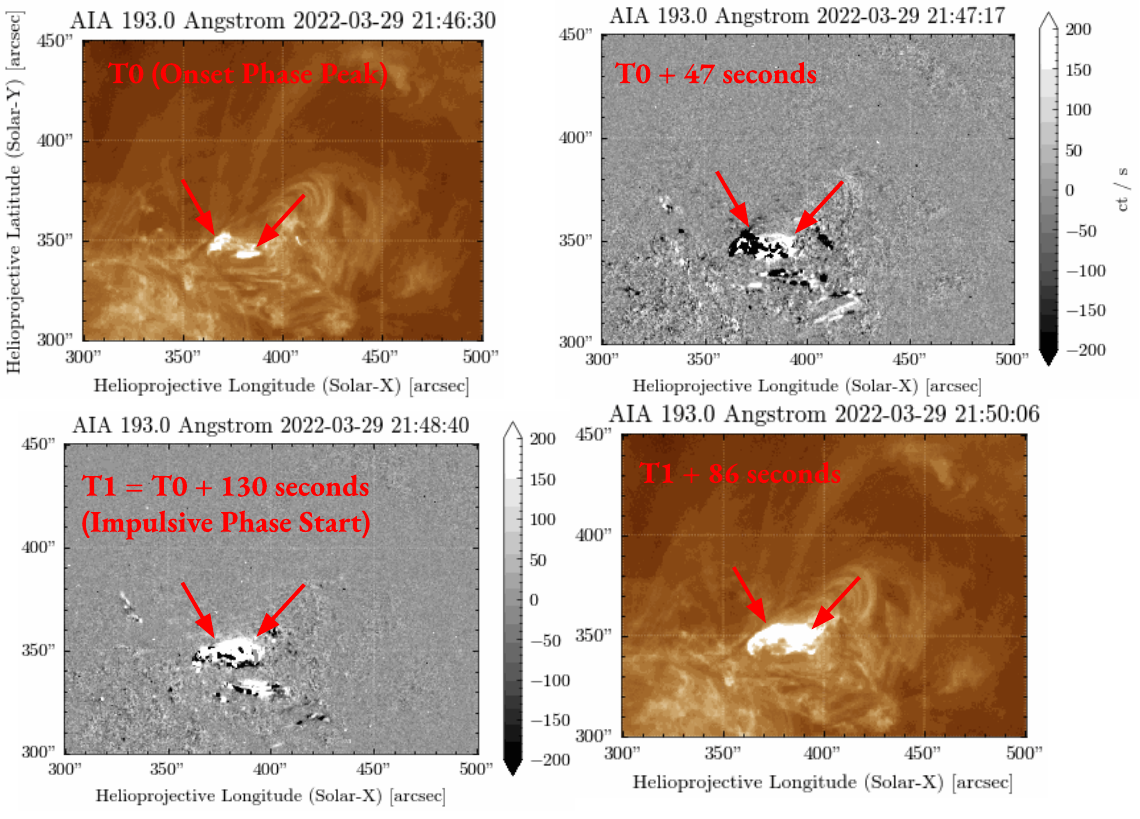}
\caption{AIA 193 \angstrom Images for a 2-loop onset Category Example: Flare-3, M1.6 flare on 2022-03-29. The first row shows images during the flare onset phase and the second row during the flare impulsive phase. The first image on the first row is at T0 (21:46:30 UTC) which is during the Soft X-ray peak of the Onset phase, and the next image is a image. The time interval of the difference image is shown in red. The first image of the second row is a running difference image at T1 (21:48:40 UTC) which is during the start of the impulsive phase. The last image is at T1 + 86 seconds (21:50:06 UTC) which is close to the end of the impulsive phase.}
\label{flare3_aia}
\end{figure}

Figure \ref{flare3_aia} shows SDO AIA 193A channel images of a M1.6 flare on 2022-03-29. The images are arranged in a similar manner as in Figure \ref{doy_066_aia}. The first image in the row shows the coronal plasma morphology during the peak of the onset phase. The next image is running difference images with the time interval after T0 labelled. The first image in the second row shows another difference image at the beginning of the impulsive phase, denoted as T1. The last image in the second row is at T1 + 86 seconds, which is close to the end of the impulsive phase. In this example, during the onset phase two small loops brighten and appear to rise or grow in size (as shown by the red arrows). These loops then interact with each other sometimes during the impulsive phase leading to brighter loops or sometimes a larger, brighter loop and subsequently theO main flare emissions. The supplementary material includes the animations for both the  AIA 193 \angstrom images (file name DOY\_108\_Col.mp4) and for the running difference images (file name DOY\_108\_Diff.mp4).

\section{Discussion and Conclusion}
Table \ref{result_table} summarizes the key results from the DAXSS soft X-ray model fitting and AIA EUV image analysis of the six solar flares. To aid discussion each flare is divided into Onset-Rising, Onset-Falling, and Main flare Impulsive phases, which are listed in the second column. The UTC time corresponding to these is listed in the third column. The hottest temperature obtained from DAXSS data fitting is listed in the fourth column. Increasing trend in temperature is denoted by $\Uparrow$, decreasing trend by $\Downarrow$, and no-trend or approximately constant is denoted by $-$. The maximum temperature for onset-rising phase, minimum temperature for onset-falling phase and maximum temperature for main-flare impulsive phase are also listed. Similarly the increasing, decreasing or constant trend of each of the elemental abundance factors are also listed in the table. The last column denotes the category of the coronal plasma morphology variation during the Onset phase of the flare. 

In all six flares, we find that the plasma temperature rises to around approximately 10 MK during the onset-rising phase. This is followed by a decrease in temperature during the onset-falling phase. Then the temperature increases again during the flare impulsive phase. This suggests that the Onset phase could have a pre-conditioning effect on the flare. The abundance factors of the Low-FIP elements for most flares also show a decrease trend during the flare onset-rising phase and main-flare impulsive phase. This highlights the important role of chromospheric evaporation during flares to inject warmer plasma into the coronal loops. Previous studies using MinXSS-1 and 2 data have shown this trend for for the main flare impulsive phase \citep{suarez_estimations_2023}. In this study we have shown that a similar trend is observed during the flare onset phase even prior to the impulsive phase.

\begin{center}
\begin{longtable}{ | m{1.3cm} | m{3cm}| m{2cm}| m{2.2cm}| m{0.5cm}| m{0.5cm}| m{0.5cm}| m{0.5cm}| m{0.5cm}| m{2.5cm}|  } 
  \hline
  \textbf{Flare No.} & \textbf{Flare Phases} & \textbf{Time (UTC)} & \textbf{Temperature (MK)} & \textbf{Mg} & \textbf{Si} & \textbf{S} & \textbf{Ca} & \textbf{Fe} &\textbf{AIA Result}\\ 
  \hline
  Flare-1 DOY065 
  & \textbf{Onset-Rising:} \newline \textbf{Onset-Falling:} \newline \textbf{Main Impulsive:}
  & 02:20 - 02:23 \newline 02:23 - 02:26 \newline 02:26 - 02:35 
  & $\Uparrow, T_{max} = 14$ \newline $\Downarrow, T_{min} = 10$ \newline $\Uparrow, T_{max} = 14$
  & $-$ \newline $-$  \newline $\Uparrow$
  & $-$ \newline $-$  \newline $-$
  & $\Downarrow$ \newline $-$  \newline $\Downarrow$ 
  & $-$ \newline $-$  \newline $\Downarrow$
  & $-$ \newline $-$  \newline $\Downarrow$
  & 1-Loop Onset
  \\ 
  \hline
  Flare-2 DOY074
  & \textbf{Onset-Rising:} \newline \textbf{Onset-Falling:} \newline \textbf{Main Impulsive:}
  & NA \newline NA \newline 23:22 - 23:26 
  & $-, T_{nom} = 9$ \newline $-, T_{nom} = 9$ \newline $\Uparrow, T_{max} = 17$
  & $-$ \newline $-$  \newline $\Downarrow$
  & $-$ \newline $-$  \newline $\Downarrow$
  & $-$ \newline $-$  \newline $\Downarrow$ 
  & $-$ \newline $-$  \newline $\Downarrow$ 
  & $-$ \newline $-$  \newline $\Downarrow$ 
  & 2-Loop Onset
  \\ 
  \hline
  Flare-3 DOY088
  & \textbf{Onset-Rising:} \newline \textbf{Onset-Falling:} \newline \textbf{Main Impulsive:}
  & 21:43 - 21:45 \newline 21:45 - 21:47 \newline 21:47 - 21:53 
  & $\Uparrow, T_{max} = 15$ \newline $\Downarrow, T_{min} = 10$ \newline $\Uparrow, T_{max} = 22.5$
  & $-$ \newline $-$  \newline $\Downarrow$
  & $-$ \newline $-$  \newline $\Downarrow$
  & $-$ \newline $-$  \newline $\Downarrow$ 
  & $-$ \newline $-$  \newline $\Downarrow$ 
  & $-$ \newline $-$  \newline $\Downarrow$  
  & 2-Loop Onset
  \\ 
  \hline
  Flare-4 DOY109
  & \textbf{Onset-Rising:} \newline \textbf{Onset-Falling:} \newline \textbf{Main Impulsive:}
  & 04:45 - 04:46 \newline 04:46 - 04:47 \newline 04:47 - 04:50 
  & $\Uparrow, T_{max} = 13$ \newline $\Downarrow, T_{min} = 11$ \newline $\Uparrow, T_{max} = 18.5$
  & $\Downarrow$ \newline $\Uparrow$  \newline $\Downarrow$
  & $\Downarrow$ \newline $\Uparrow$  \newline $\Downarrow$
  & $-$ \newline $\Uparrow$  \newline $\Downarrow$ 
  & $\Downarrow$ \newline $\Uparrow$  \newline $\Downarrow$ 
  & $\Downarrow$ \newline $\Uparrow$  \newline $\Downarrow$ 
  & 1-Loop Onset
  \\ 
  \hline
  Flare-5 DOY227
  & \textbf{Onset-Rising:} \newline \textbf{Onset-Falling:} \newline \textbf{Main Impulsive:}
  & 17:22 - 17:25 \newline 17:25 - 17:30 \newline 17:30 - 17:37 
  & $\Uparrow, T_{max} = 9$ \newline $\Downarrow, T_{min} = 7$ \newline $\Uparrow, T_{max} = 20$
  & $-$ \newline $-$  \newline $\Downarrow$
  & $-$ \newline $-$  \newline $\Downarrow$
  & $-$ \newline $-$  \newline $\Downarrow$ 
  & $-$ \newline $-$  \newline $\Downarrow$ 
  & $-$ \newline $-$  \newline $\Downarrow$ 
  & 2-Loop Onset
  \\ 
  \hline
  Flare-6 DOY230
  & \textbf{Onset-Rising:} \newline \textbf{Onset-Falling:} \newline \textbf{Main Impulsive:}
  & 10:20 - 10:31 \newline 10:31 - 10:38 \newline 10:38 - 10:55
  & $\Uparrow, T_{max} = 12.5$ \newline $\Downarrow, T_{min} = 11.5$ \newline $\Uparrow, T_{min} = 19$
  & $\Downarrow$ \newline $\Uparrow$  \newline $\Downarrow$
  & $\Downarrow$ \newline $\Uparrow$  \newline $\Downarrow$
  & $\Downarrow$ \newline $\Uparrow$  \newline $\Downarrow$ 
  & $\Downarrow$ \newline $\Uparrow$  \newline $\Downarrow$ 
  & $\Downarrow$ \newline $\Uparrow$  \newline $\Downarrow$ 
  & 2-loop Onset
  \\ 
  \hline
\caption{Key model fit results are summarized for the onset-rising peak time, onset-falling (valley) time and flare main flare impulsive peak time. For each of these times the hottest temperature (Temp-3) obtained from model fitting is listed. The trend in variation of hottest temperature and abundance factors during these flare phases are also listed. Increasing trend in temperature is denoted by $\Uparrow$, decreasing trend by $\Downarrow$, and no-trend or approximately constant is denoted by $-$.}
\label{result_table}
\end{longtable}
\end{center}

The AIA image analysis reveal two categories of morphology variations both of which show interaction coronal loops during the onset phase, which eventually trigger the main impulsive phase of the flare.  The first 1-loop onset category comprises of a small loop that brightens during the Onset, but it is distinct from the larger loop that brightens during the main flare. The onset smaller loop interacts with the larger (assumed higher) loop through the smaller loop heating and rising up to the larger loop or through loop footpoint interactions. Although two loops are involved, only one loop dominates at a time for the 1-loop onset category. The second category is when two loops emerge or brighten during the onset, and they then interact and sometimes merge together to form brighter loops and sometimes a single larger, brighter loop during the impulsive phase. For this 2-loop onset category, two loops brighten at the same time. These two geometries are observed to be similar to the flare geometries highlighted in \cite{benz_flare_2016_updated} and references therein. The 1-loop geometry corresponds to the standard CSHKP model, and the 2-loop scenario comprises of two different loops merging together to form the main loop. It should be noted that these are not completely mutually exclusive categories, i.e. flares in both categories can also exhibit some hybrid properties (a combination of the two) and there could be are other possibilities of flare initiation. One key conclusion is that the onset flare physics appears to be very similar to the main flare physics, but with the onset phase being dimmer than the main impulsive phase. This is highlighted both by the DAXSS spectrum fit results as well as the AIA image analysis.

\section{Future Work}
Our next step in the investigation of X-ray onsets will be to combine DAXSS soft X-ray spectra with other Hard X-ray measurements at higher energies. Instruments such as Solar Orbiter Spectrometer/Telescope for Imaging X-rays \citep{stix} would be suitable for this analysis. Additionally, it will also be useful to cross-validate DAXSS modeling results with other Soft X-ray spectrometers. The Chandrayaan-2 Solar X-ray Monitor (XSM) \citep{xsm} and Solar Orbiter Spectrometer/Telescope for Imaging X-rays  \citep{stix} instruments with similar measurement capabilities was also in operation in 2022, which allows planning of model cross validations and instrument comparison campaigns together. Such efforts have already begun with one of the flares in this paper being analyzed with XSM measurements. Lastly, to further understand the plasma morphology variation during the onset phase, multispectal EUV images can also be analysed along with Magnetogram images to characterize the changes in magnetic configurations during flares. Our long-term goal is understand the flare onset physics better and if onset behavior could be a predictive indicator for the main flare magnitude and timing.

\section{Data Source}
The DAXSS solar SXR spectral irradiance data products, along with the DAXSS RMF and ARF calibration files, user guide, and data plotting examples in IDL and Python, are available from the MinXSS website at
\url{http://lasp.colorado.edu/home/minxss/}. The GOES XRS data are available from
\url{http://www.ngdc.noaa.gov/stp/satellite/goes/dataaccess.html}. The SDO-AIA images are courtesy of NASA/SDO and the AIA and HMI science teams, and have been accessed from \url{http://jsoc.stanford.edu/}.

\bibliography{references,sample631}{}
\bibliographystyle{aasjournal}

\end{document}